# Correlation between Thermal Properties, Electrical Conductivity and Crystal Structure in the BaCe$_{0.80}$Y$_{0.20}$O$_{2.9}$ Proton Conductor

*Lorenzo Malavasi[1,2*], Clemens Ritter[3], Gaetano Chiodelli[2]*

[1]Dipartimento di Chimica Fisica and [2]CNR - IENI Sezione di Pavia - Università di Pavia

[3]Institute Laue-Langevin, Boite Postale 156, F-38042, Grenoble, France.

**ABSTRACT**

In this paper we report an extensive neutron diffraction investigation at high temperature on the BaCe$_{0.80}$Y$_{0.20}$O$_{2.9}$ proton conducting material. Our results precisely define the structural evolution of the compound as a function of temperature which is from a monoclinic (room temperature) to a cubic (800°C) structure. Neutron data have been correlated to calorimetric measurements (TGA and DSC) and conductivity properties of the material.

*Corresponding Author: Dr. Lorenzo Malavasi, CNR - IENI Sezione di Pavia and Dipartimento di Chimica Fisica-Università di Pavia, V.le Taramelli 16, I-27100, Pavia, Italy. Tel: +39-0382-987921 - Fax: +39-0382-987575 - E-mail: lorenzo.malavasi@unipv.it

# Introduction

Many efforts in the field of solid state ionics have focused in the last two decades on the study of perovskite oxides exhibiting significant proton conductivity due to their possible application in solid state electrochemical devices such as fuel cells, gas sensors, and hydrogen pumps [1-9]. With the progress in the synthesis and characterization of the newly discovered materials, such as the $BaCeO_3$ and $CaZrO_3$ perovskites, an intense research has also been devoted to the comprehension of the defect chemistry of the same materials, in order to optimize their performance and build the basis for the development of other solid-state proton conductors [10-15], which are recently attractive candidates for applications in intermediate-temperature solid oxide fuel cells (IT-SOFCs). In fact, proton conducting electrolytes for IT-SOFCs have also the advantage of water production at the cathode, thus avoiding the fuel dilution at the anode.

One of these perovskites showing promising properties in the field of solid state proton conductors is the aliovalent-doped barium cerate, $BaCe_{1-x}A_xO_{3-\delta}$ (A=commonly rare earth, $\delta=x/2$). $BaCeO_3$ has been reported to have a total conductivity of about $5.3\times10^{-2}$ $\Omega^{-1}cm^{-1}$ when it is doped with 20% Y ($BaCe_{0.80}Y_{0.20}O_{2.9}$) [14]. Guan and co-workers [16] provided evidence for increasing total conductivity in $BaCe_{1-x}Y_xO_{3-\delta}$ with the increase of Y concentration up to $x=0.2$. The increase in the proton conductivity with doping is mainly related to the creation of oxygen vacancies through the defect reaction (written in Kroger-Vink notation):

$$2Ce_{Ce}^{X} + Y_2O_3 + O_O^{X} \rightarrow 2Y_{Ce}^{'} + V_O^{\bullet\bullet} + 2CeO_2 \tag{1}$$

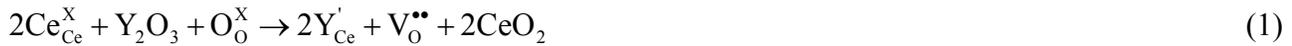

with oxygen vacancies constituting the site for the incorporation of water in the form of hydroxyl groups through:

$$H_2O_g + V_O^{\bullet\bullet} + O_O^{X} \Leftrightarrow 2OH_O^{\bullet} \tag{2}$$

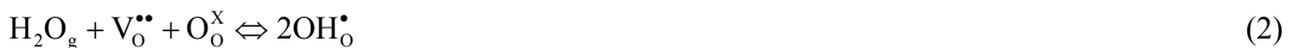

Whereas oxygen vacancies represent the main defects at high temperatures where water molecule desorption and oxygen migration take place in the bulk, the dissolution of protons is favoured by decreasing temperatures. In fact, proton incorporation is found to be exothermic, suggesting an energetic stabilisation of the protonic defects with increased doping.[17,18]

An extremely relevant aspect of any material that should be used in a device is the study of its structural behaviour as a function of temperature. In the present literature it is possible to find exhaustive work related to the high-temperature structural behaviour of pure $BaCeO_3$ [19,20]. The seminal neutron diffraction work by Knight precisely defined a quite complex trend in the structural evolution of $BaCeO_3$; it was found that this material undergoes three structural phase transitions between 4.2 and 1273 K: a) from 4.2 to 563 K, it is orthorhombic, with space group *Pnma*; b) at 563 K, it undergoes a second order phase transition to a second orthorhombic phase with space group *Immm*; c) at 673 K, it undergoes a first order phase transition to a rhombohedral structure with space group $R\bar{3}c$; d) at 1173 K, it transforms to the a cubic phase with space group $Pm\bar{3}m$. The same paper discusses as well the structural properties of the $BaCe_{0.90}Y_{0.10}O_{2.95}$ sample which shows an analogous trend in the crystal structure evolution with temperature as $BaCeO_3$.

The effect of the Y-concentration in $BaCe_{1-x}Y_xO_{3-\delta}$ solid solution for $0 \leq x \leq 0.3$ is considered in response to heat treatments in oxidising, reducing and water-vapour containing atmospheres [21]. However, the neutron diffraction data reported refer only to room-temperature patterns.

To the best of our knowledge the only paper reporting some high-temperature neutron data compassing also the $BaCe_{0.80}Y_{0.20}O_{2.9}$ material is a recent work by C.-K. Loong *et al*. [22]. This paper provides some information about the evolution with *T* of the $BaCe_{1-x}Y_xO_{3-\delta}$ solid solution for $0 \leq x \leq 0.3$ but it is mainly devoted to the description of the application of small-to-wide angle diffraction and inelastic scattering to the study of this system without providing a detailed investigation of the high temperature structural evolution and features of $BaCe_{0.80}Y_{0.20}O_{2.9}$.

In view of these considerations we believe that a detailed and through analysis of the optimal doped $BaCe_{0.80}Y_{0.20}O_{2.9}$ proton conductor is highly demanded. For this purpose we carried out a neutron diffraction investigation of BCY as a function of temperature between room temperature (RT) and 800°C. The structural investigation has been completed by means of thermal analysis (thermogravimetry and differential scanning calorimetry) and electrical conductivity measurements.

# Experimental Section

Powder samples of $BaCe_{0.80}Y_{0.20}O_{2.9}$ have been prepared by conventional solid state reaction from the proper stoichiometric amounts of BaO, $CeO_2$, and $Y_2O_3$ (all Aldrich ≥ 99,9%) by two successive firings for 20 h at 1200 °C.

The chemical composition was checked by means of micro-probe analysis which revealed a sample composition in agreement with the nominal one.

Neutron powder diffraction data were acquired on the D1A instrument at the Institute Laue Langevin (ILL) in Grenoble. All measurements were recorded in air in a silica glass containe. The diffraction patterns were collected in the angular range 0°-160°, step 0.05° at a wavelength 1.39 Å, for about 6 hours per spectrum in the temperature range between room temperature and 800°C every 100°C.

All the neutron diffraction patterns were analysed according to the Rietveld method [23,24] by means of the FULLPROF software package [25]. For the XRPD patterns the background was fitted with an interpolation between fixed point chosen outside Bragg peaks. Cell parameters, atomic position and isotropic thermal factors for all the ions were refined. For the NPD patterns the background coming from the empty quartz tube was recorded at the same temperatures and subtracted from the NPD patterns of the samples. Typical $R_{wp}$ values for the NPD (XRPD) refinements are around 7-12 (8-12) with $\chi^2$ values between 1.2 and 2 (1.5-2) for all the temperatures and samples.

AC electrical characterizations were made by using the Impedance Spectroscopy (IS) technique with a Frequency Response Analyser (FRA) Amel 7200 or Solatron 1260 apparatus, in the frequency range $10^{-3}$–$10^7$ Hz. A home made high-impedance adapter (up to $10^{13}$ ohm, 2 pF) with an active guard driver was applied to the FRA in order to reduce the noise and the parasitic capacitances of cables and cells and to increase the sensitivity and input impedance of the apparatus. High temperature measurements (up to 900°C) were performed by inserting the electrochemical cell in an oven. In order to ensure good electrical contact for AC measurements, platinum layers were deposited by sputtering on the flat surfaces of the pellets.

Thermogravimetry (TGA) measurements were carried out under the same conditions as the diffraction measurements (i.e. in air) from RT to 800°C with a heating rate of 5°C/min with a TA2950 instrument. Differential scanning calorimetry (DSC) were carried out in air on the powdered samples using a SDT2960 TA instrument in the range 25-800°C at a heating rate of 5°C/min.

# 3. Results and Discussion

**3.1 Crystal Structure Evolution of BCY with *T***

Figure 1 displays the neutron diffraction (ND) patterns acquired on the $BaCe_{0.80}Y_{0.20}O_{2.9}$ sample at the different temperatures investigated (30-800°C). The inset highlights a region showing clear changes. As already previously stated, neutron diffraction is a more sensitive probe with respect to X-ray diffraction for the study of cerates due to both the possibility to get valuable information related to superlattice intensity and for the lack of form-factor fall with $Q$ [20].

The simple visual inspection of the pattern changes with temperature suggests an increase of symmetry by increasing temperature and, in addition, that the structural changes occur gradually with temperature. Just to make some representative examples, the main peak, located at about 36°, is found to be splitted at RT and progressively becomes a single peak at 600°C; in the same way the three peaks between 38 and 42° first merge in 2 peaks at 500C° and then just one peak is left at 800°C.

The ND data have been analysed by means of Rietveld refinement strategy. However, for all the temperatures, we firstly checked the validity of the proposed crystal structures by means of a matching-profile fit. For the data at RT we found that the correct description of the observed pattern is consistent with a monoclinic unit cell belonging to the *I*2/*m* space group (No. 12). In addition, about 7% of a secondary phase has been detected. The nature of this secondary phase is not easy to be assessed since for both rhombohedral and orthorhombic phases (both *Imma* and *Pnma*) the main peaks are located more or less at the same position. The best $\chi^2$ (5.4) was obtained by considering an orthorhombic *Imma* secondary phase with respect to other phases ($\chi^2>6$). The final refinement of the RT ND patter is shown in Figure 2. Structural data for the monoclinic phase are reported in Table 1.

The monoclinic structure has been found also at 300 and 400°C. In Figure 3 is reported the trend of the pseudo-cubic cell volume while in Figure 4 is reported the behaviour of the lattice parameters as a function of temperature. As can be appreciated the cell volume expands linearly with temperature. The lattice constants reported in the Figure have been expressed in terms of a pseudo-cubic cell, neglecting the small monoclinic distortion ($\beta$ angle $\neq$ 90). The three lattice parameters tend to get closer as the temperature increases and this is particularly evident for the *a* and *c* lattice constants. The evolution of the monoclinic $\beta$ angle with temperature is reported in Figure 5. As the temperature increases the distortion of the unit cell is reduced as witnessed by the $\beta$ angle moving progressively closer to 90°.

At 500°C the BCY neutron pattern can not be modelled anymore with a monoclinic structure. In particular, in the region of the main peak (around 36°) the previously observed splitting of the (-202) and (202), due to the deviation of the $\beta$ angle from 90°, has disappeared and a single symmetric peak is now present. Several other examples of splitting removal are present in the pattern.

According to the only available high-temperature structural study on $BaCe_{0.80}Y_{0.20}O_{2.9}$ [22] at 500°C the structure should be, mainly, in the *Pnma* orthorhombic phase (under oxidative atmosphere). However, this crystal structure is not compatible with our diffraction pattern. Figure 6 reports a simulation of the expected diffraction pattern in the case of a *Pnma* and in the case of a *Imma* orthorhombic structure. As can be appreciated, the experimental pattern at 500°C does not display those peculiar reflections of the *Pnma* symmetry which are partially (only the most intense ones) highlighted in the bottom part of Figure 6 with arrows. The pattern at 500°C has been also checked with respect to the rhombohedral $R\bar{3}c$ symmetry. The presence of some peak splitting in the experimental ND pattern at high angles (particularly above 75°) together with some peak broadening clearly rule out the presence of the rhombohedral symmetry at this temperature. The collected evidence totally support the formation of an *Imma* orthorhombic phase at 500°C. We finally note that the *Imma* space group and the *I2/m* monoclinic space group are correlated by a group-subgroup relation. The quality of the refinement of neutron data at this temperature with the proposed structure can be inferred by looking at Figure 7.

The pseudo-cubic cell volume, the lattice parameters and the refinement results are reported, respectively, in Figure 3, 4 and Table 1. From Figure 3 we note that there are no detectable discontinuities in the cell volume trend at the *I2/m→Imma* phase transition. This suggest a second order phase transition which is in agreement with a group-subgroup relationship between the two phases and with the continuous and progressive reduction of structure distortion from RT to 500°C. In addition, no clear calorimetric peaks are found in correspondence with this transition (see later in the text).

The further raise of temperature to 600°C leads to the appearance of a new crystal structure. This is well witnessed by the reduction of peak broadening and removal of several peak splittings which is particularly evident at high angle (see inset of Figure 7b). The new crystal structure adopted by the $BaCe_{0.80}Y_{0.20}O_{2.9}$ sample is the $R\bar{3}c$ rhombohedral structure. The refinement of the ND experimental pattern at this temperature is shown in the main part of Figure 7b. The pseudo-cubic cell volume, the lattice parameters and the refinement results at 600°C are reported, respectively, in Figures 3, 4 and Table 1. No sign of any secondary phase is found. The *Imma→$R\bar{3}c$* phase transition does not occur with any evident discontinuity in the pseudo-cubic cell volume. This is in contradiction with a previous report by Knight on the $BaCeO_3$ material [19] in which it was shown a significant discontinuity in the cell volume trend of about 0.5% at this phase transition. However, our results are in agreement with those reported in a neutron diffraction study on the $BaCe_xZn_{1-x}O_3$ solid solution [26] and in agreement with a previous work on the Raman spectroscopy properties of pure $BaCeO_3$ [27]. We may note, however, that, to the best of our knowledge, this is the first paper reporting the structural details, as a function of temperature, for the $BaCe_{0.80}Y_{0.20}O_{2.9}$ sample and that significant differences may be present with respect to the pure compound. For the 10% Y-doping Knight found the same sequence of phase transitions of the pure samples but with modified widths of the phase field [20]. The further increase of Y-doping leads to a different sequence of structural changes without the observed discontinuity in the cell volume trend found previously.

The rhombohedral structure is also found at 700°C. Again, the pseudo-cubic cell volume, the lattice parameters and the refinement results at this temperature are reported, respectively, in Figure 3, 4 and Table 1.

As can be appreciated from Figure 1, at 800°C the BCY structure undergoes a further symmetrisation as clearly evidenced by the reduction of Bragg peaks found in the pattern. The crystal structure adopted by the sample at this temperature is the cubic $Pm\bar{3}m$ structure. This result agrees with the results found by Knight, where for the pure $BaCeO_3$ the cubic structure is observed at 900°C [20]. The Rietveld refined pattern is shown in Figure 7c. Beside the cubic phase a second rhombohedral phase has been added in the refinement. The amount of the latter, according to the refinement results, was 1.2(2)%. As in the previous cases, the cubic cell volume, the lattice parameter and the refinement results at 800°C are reported, respectively, in Figure 3, 4 and Table 1. As for the previous cases, we did not observe any strong discontinuity in the unit cell volume at the $R\bar{3}c \rightarrow Pm\bar{3}m$ phase transition although a deviation from the linear behaviour with $T$ observed for the data up to this temperature can be inferred at 800°C from Figure 3. The origin of this slight deviation is not easily assessed since the $BaCe_{0.80}Y_{0.20}O_{2.9}$ compound undergoes oxygen content loss with increasing temperature (particularly above 700°C, see later in the text) as well as a progressive change in the Ce oxidation state which may lead to a non-linear trend in the $V$ vs. $T$ data. From the structural and calorimetric data it is possible to conclude that this phase transition is of second order.

## 3.2 Evolution of the Structural Parameters with $T$

From the Rietveld refinements we calculated the bond lengths and bond angles at the different temperatures explored.

Figure 8 displays the average Ba-O and B-O bond lengths (B=Ce and Y) as a function of temperature for $BaCe_{0.80}Y_{0.20}O_{2.9}$. Regarding the B-O bond there are three distinct lengths in the monoclinic structure, two in the orthorhombic and one in the rhombohedral and cubic crystal structures. It can be appreciated that the temperature evolution of the B-O and Ba-O bond lengths is opposite: in the first case there is a progressive contraction of the bond as the temperature increases with a severe contraction at 800°C, while in the second case the bond length expands up to 700°C and then slightly shrinks from 700 to 800°C.

Based on the average bond lengths derived from the Rietveld refinements we calculated the geometrical tolerance factor ($t$):

$$t = \frac{\langle Ba-O \rangle}{\sqrt{2}\langle B-O \rangle} \tag{1}$$

which is a measure of the distortion degree of a perovskite compound. The results are summarized in the inset of Figure 8. The value of $t$ progressively increases from RT to 800°C as the perovskite structure becomes more and more regular. As the structure gets cubic, as expected, the tolerance factor becomes 1. The values of the bond lengths are reported in Table 2.

The temperature evolution of the average B-O-B bond angle and the <Ba-B> bond length are reported in Figure 9. In agreement with the increase of the symmetry of the perovskite structure the bond angle moves from about 155° to 180° (in the cubic structure). Let us remember that in the monoclinic and orthorhombic structures there are two distinct B-O-B bond angles (one within the equatorial plane and the other in the axial plane). The values of the two angles are reported in Table 2 together with the Ba-B average bond length. In addition, as the B-O environment becomes more compact by reduction of the bond length the distance between the cations increases linearly with temperature by about 0.8%.

Figure 10 shows the temperature variation of the atomic displacement parameters (a.d.p.) for the BCY compound. For the oxygen ions we modelled the a.d.p. by means of an anisotropic thermal factor. In the Figure reported is the resulting isotropic parameter ($B$). The overall behaviour shows a progressive increase of all the $B$ parameters as the temperature increases without any significant anomaly except for a quite significant rise of the O a.d.p. at 800°C. This can be correlated to the high mobility of oxygen ions at high temperature in the $BaCe_{0.80}Y_{0.20}O_{2.9}$ material.

Finally, Figure 11 reports the variation of the oxygen content as derived from the refinement of the oxygen atoms occupancies. In the Figure we also plotted a thermogravimetric trace of $BaCe_{0.80}Y_{0.20}O_{2.9}$ collected under the same experimental conditions employed for the acquisition of the neutron diffraction patterns. First of all we note that the refined occupancy at RT is around 2.90(4), in good agreement with the nominal oxygen content of the sample. This oxygen content remains practically constant until about 600°C and then starts to decrease. The oxygen content derived from oxygen occupancy at 800°C is 2.79(4). The comparison with the TGA trace reveals a good overlap between the weight variation and the oxygen content change with temperature. The weight change in the thermogravimetric measure is ascribable to the change in the oxygen content only. This can be concluded by the reversible behaviour of the weight variation with *T* (not shown). The oxygen content at 800°C determined from the TGA curve is 2.82, which is in fairly good agreement with the oxygen content determined from the Rietveld refinement.

Regarding the localization of the oxygen vacancies, we could observe that the O1 site in the monoclinic structure, *i.e.* the apical oxygen, resulted to be always fully occupied within the experimental error, while the O2 and O3 sites, *i.e.* the equatorial oxygens, showed partial occupancies, particularly in the O2 site. A sketch of the monoclinic structure with the atom labels is reported in Figure 12a. Even though there are no previous studies on monoclinic phases of doped barium cerates, the result of oxygen vacancies mainly located on the equatorial plane is in good agreement with a similar work by Kruth *et al*. on the orthorhombic $Ba_{1-x}La_xCe_{0.9-x}Y_{0.1+x}O_{2.95}$ solid solution [28] where, by means of combined experimental and computational techniques, it was found that the preferential site for the oxygen vacancy location (and consequently of water incorporation) mainly involves the O2 site (*i.e.* the equatorial oxygen of the orthorhombic crystal structure). Also in the present case, for the orthorhombic phase of $BaCe_{0.80}Y_{0.20}O_{2.9}$ the oxygen occupancy was found to be complete for the O1 site and partial for the O2 site. Figure 12b reports a sketch of the *Imma* structure found at 500°C for the $BaCe_{0.80}Y_{0.20}O_{2.9}$ sample. For sake of completeness we also plotted in Figure 12c and 12d the rhombohedral and cubic structures, respectively.

## 3.3 Correlation with Transport Properties

Figure 13 displays the Arrhenius plot of the electrical conductivity data of the BaCe$_{0.80}$Y$_{0.20}$O$_{2.9}$ sample collected under the same conditions as the neutron diffraction experiments. The conductivity increases as the temperature increases with an evident slope change around 500°C. The activation energies before and after the slope change are about 0.85 eV and 0.35 eV, respectively. Under the measurement conditions the only mobile defects are oxygen vacancies. Their concentration remains practically constant until 700°C as determined in this work by means of TGA and Rietveld refinement of oxygen occupancies (see Figure 11). The increase of conductivity by raising temperature is due to an increase of defects mobility until the highest temperatures, where the reduction in the oxygen content contributes to the increase in the oxygen vacancies concentration. The observed change in the activation energy falls in the region of a phase transition of BaCe$_{0.80}$Y$_{0.20}$O$_{2.9}$ (*Imma*→$R\bar{3}c$). The trend in the conductivity data coupled to the structural evidences suggests that this transition is of second order in nature. This is also supported by the calorimetric measure we have carried out on the sample again under the same conditions of ND and electrical transport measurements.

The differential scanning calorimeter (DSC) trace of BaCe$_{0.80}$Y$_{0.20}$O$_{2.9}$ is reported in the inset of Figure 13. As can be appreciated there are no clear peaks witnessing an exo- or endothermic thermal event associated with any structural phase transition. This allows us to conclude that the slope change observed in the electrical conductivity data is associated with an increase in the perovskite structure symmetry which is coupled to the *Imma*→$R\bar{3}c$ transition.

# Conclusions

In this work we aimed at presenting the structural evolution with temperature of the BaCe$_{0.80}$Y$_{0.20}$O$_{2.9}$ proton conducting oxide. To the best of our knowledge this is the first high-temperature investigation on this stoichiometry which represents the one displaying the optimal conductivity performance [2]. Previous completed ND works concentrated on pure BaCeO$_3$ [20,29] while the only available paper on BaCe$_{0.80}$Y$_{0.20}$O$_{2.9}$ did not report any detailed structural insight into the temperature evolution of the material [21].

The structural results allowed us to define the evolution of the crystal structure of BaCe$_{0.80}$Y$_{0.20}$O$_{2.9}$ as a function of temperature. It can be summarized in the following:

a) from RT to 400 it is monoclinic with space group *I2/m*;

b) at 500°C the materials adopts the orthorhombic *Imma* structure;

c) at 600 and 700°C it is rhombohedral ($R\bar{3}c$) and

d) finally at 800°C it becomes cubic ($Pm\bar{3}m$).

As the temperature increases the octahedral tilting progressively reduces. In the monoclinic structure we are facing a two-tilt system known, in the Glazer's notation [30] as $a°b^-c^-$. The *I2/m→Imma* phase transition makes the tilt around the *z*-axis the same as around the *y*-axes since it can be written as a $a°b^-c^- \rightarrow a°b^-b^-$ transition. Again, we are in a two-tilt system. Further increase of temperature to 600°C removes one tilting bringing the symmetry of the system to the $a^-a^-a^-$ tilt system, *i.e.* the rotation angle is the same about each of the three axes with rotations of two neighbouring octahedra, along the tilt axis, which are in opposite directions. Finally, the transition to the cubic $Pm\bar{3}m$ space group is described as a $a°a°a°$ tilt system which excludes any tilting between octahedra.

This progressive increase in the symmetry of the system is not accompanied by any first order phase transition. This can be concluded from the thermal analysis carried out on the sample under the same conditions as the ND acquisition (see Figure 13) and is supported by the absence of any relevant discontinuity in the cell volume evolution with temperature. A slight deviation of the cell volume at high temperature is most probably related to the variation of the oxygen content above 700°C (see Figure 11). Our data allowed us to correlate the slope change in the electrical conductivity data to the second order $Imma \rightarrow R\bar{3}c$ phase transition.

Combined neutron data and TGA measurement shows that the oxygen content in the sample is stable up to about 700°C. Above this temperature the oxygen content reduces from the nominal content of about 2.9 to about 2.8. The oxygen vacancies, which are the preferential sites for the proton incorporation within the $BaCe_{0.80}Y_{0.20}O_{2.9}$ material are preferentially located in the equatorial positions of the $BO_6$ octahedron. In addition, for the monoclinic structure, the O2 site seems to be the preferential site for the localization of oxygen vacancies. Our results provide direct evidence that the $BaCe_{0.80}Y_{0.20}O_{2.9}$ proton conducting oxide is a stable material without any volume discontinuity within the operational range of an intermediate fuel cell which is a fundamental pre-requisite for the incorporation of an electrolyte within a device.

# Acknowledgement

This work has been supported by the "Celle a combustibile ad elettroliti polimerici e ceramici: dimostrazione di sistemi e sviluppo di nuovi materiali" FISR Project of Italian MIUR. We recognize the support of the UNIPV-Regione Lombardia Project on Material Science and Biomedicine. ILLL neutron facility and European Community financial support is acknowledged.

**Table 1** Structural data for $BaCe_{0.80}Y_{0.20}O_{2.9}$ at the different temperatures investigated.

|  |  | RT | 300°C | 400°C | 500°C | 600°C | 700°C | 800°C |
|---|---|---|---|---|---|---|---|---|
| Space Group |  | $I2/m$ | $I2/m$ | $I2/m$ | $Imma$ | $R\bar{3}c$ | $R\bar{3}c$ | $Pm\bar{3}m$ |
| a (Å) |  | 6.2487(1) | 6.2579(1) | 6.2589(1) | 6.2568(6) | 6.26424(6) | 6.27138(6) | 4.4359(2) |
| b (Å) |  | 8.7419(1) | 8.7817(1) | 8.8037(2) | 8.8268(7) | 6.26424(6) | 6.27138(6) | 4.4359(2) |
| c (Å) |  | 6.2339(1) | 6.2538(1) | 6.2588(1) | 6.2711(6) | 15.34041(2) | 15.35102(2) | 4.4359(2) |
| β (°) |  | 90.9962(9) | 90.8489(9) | 90.498(1) |  |  |  |  |
| V (Å³) |  | 340.50(8) | 343.64(8) | 344.8(1) | 346.34(6) | 521.3(1) | 522.8(1) | 87.288(9) |
| Ba | x | 0.262(1) | 0.256(2) | 0.252(3) | 0 | 0 | 0 | 0.5 |
|  | y | 0 | 0 | 0 | 0.25 | 0 | 0 | 0.5 |
|  | z | 0.737(1) | 0.736(2) | 0.736(3) | 0.995(2) | 0.25 | 0.25 | 0.5 |
|  | B | 1.5(1) | 2.1(1) | 1.9(1) | 2.4(1) | 2.7(1) | 3.0(1) | 4.0(2) |
| Ce/Y | x | 0.25 | 0.25 | 0.25 | 0 | 0 | 0 | 0 |
|  | y | 0.25 | 0.25 | 0.25 | 0 | 0 | 0 | 0 |
|  | z | 0.25 | 0.25 | 0.25 | 0.5 | 0 | 0 | 0 |
|  | B | 0.47(9) | 0.7(1) | 1.0(1) | 0.77(8) | 1.11(8) | 1.30(9) | 1.7(1) |
| O1 | x | 0.183(1) | 0.193(2) | 0.194(2) | 0 | 0.4515(4) | 0.4573(6) | 0 |
|  | y | 0 | 0 | 0 | 0.25 | 0 | 0 | 0.5 |
|  | z | 0.228(1) | 0.225(2) | 0.239(3) | 0.435(1) | 0.25 | 0.25 | 0 |
|  | B | 2.35(9) | 4.3(1) | 4.5(1) | 3.0(1) | 3.6(2) | 4.2(2) | 6.5(3) |
| O2 | x | 0 | 0 | 0 | 0.25 |  |  |  |
|  | y | 0.310(1) | 0.302(2) | 0.298(2) | 0.5317(7) |  |  |  |
|  | z | 0 | 0 | 0 | 0.25 |  |  |  |
|  | B | 2.5(1) | 2.9(1) | 2.8(2) | 3.4(1) |  |  |  |
| O3 | x | 0.5 | 0.5 | 0.5 |  |  |  |  |
|  | y | 0.2241(9) | 0.226(1) | 0.225(2) |  |  |  |  |
|  | z | 0 | 0 | 0 |  |  |  |  |
|  | B | 2.0(1) | 3.0(1) | 3.6(2) |  |  |  |  |

Table 2 Bond lengths and angles for $BaCe_{0.80}Y_{0.20}O_{2.9}$ at the different temperatures investigated.

| | RT | 300°C | 400°C | 500°C | 600°C | 700°C | 800°C |
|---|---|---|---|---|---|---|---|
| Parameter | $I2/m$ | $I2/m$ | $I2/m$ | $Imma$ | $R\bar{3}c$ | $R\bar{3}c$ | $Pm\bar{3}m$ |
| B-O1 (Å) | 2.228(2) | 2.229(2) | 2.230(2) | 2.244(2) | 2.235(4) | 2.233(5) | 2.218(1) |
| B-O2 (Å) | 2.251(3) | 2.242(3) | 2.243(3) | 2.232(5) | | | |
| B-O3 (Å) | 2.237(1) | 2.238(1) | 2.233(2) | | | | |
| B-O1-B (°) | 157.49(7) | 159.99(9) | 161.59(7) | 159.13(6) | 164.40(1) | 166.25(1) | 180 |
| B-O2-B (°) | 152.82(8) | 156.50(8) | 158.30(8) | 165.60(3) | | | |
| <Ba-O> (Å) | 3.125(2) | 3.122(2) | 3.1334(2) | 3.137(2) | 3.139(2) | 3.141(3) | 3.137(1) |
| <Ba-Ce> (Å) | 3.811(1) | 3.822(1) | 3.826(1) | 3.829(2) | 3.836(1) | 3.838(1) | 3.842(1) |
| t | 0.985 | 0.989 | 0.990 | 0.992 | 0.993 | 0.995 | 1 |
| $R_{wp}$ | 6.35 | 7.1 | 7.9 | 7.2 | 7.2 | 7.8 | 7.1 |
| $\chi^2$ | 4.05 | 4.9 | 5.1 | 4.5 | 3.6 | 4.6 | 3.9 |

# Figures

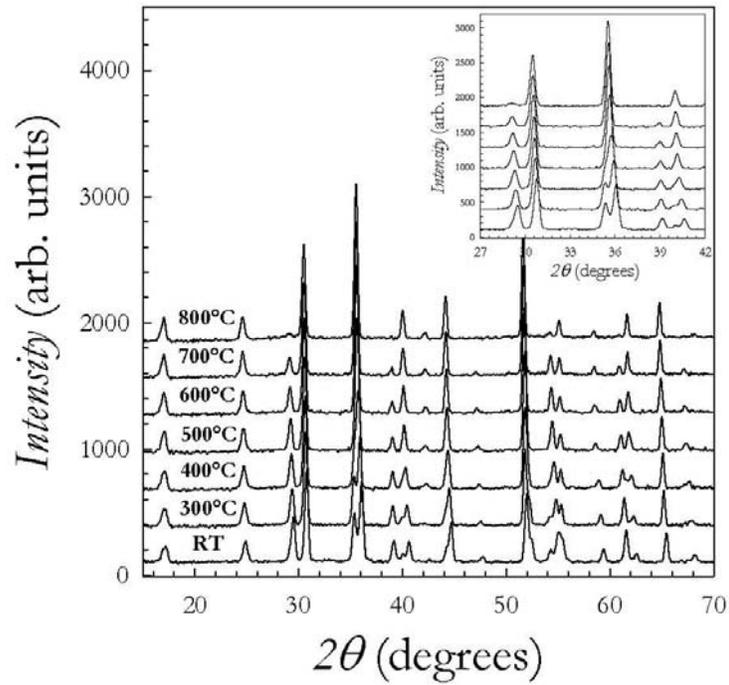

**Figure 1** – Neutron Diffraction (ND) patterns of BCY at the temperatures investigated. Inset: details of ND patterns around the region where main peaks are located. Temperatures are the same as in the main Figure.

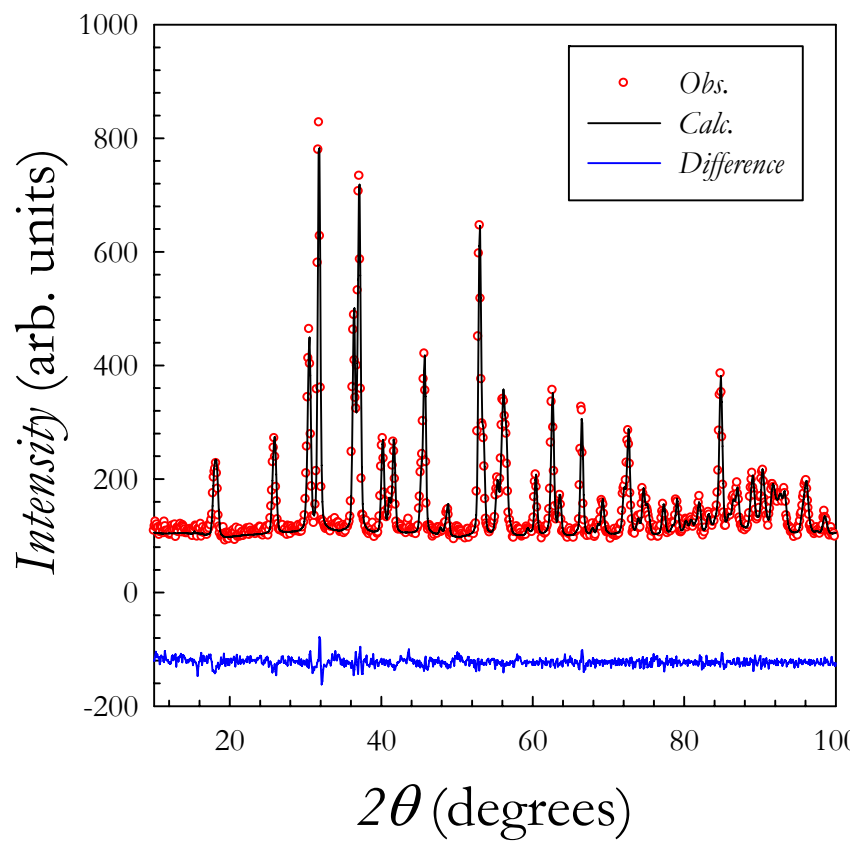

**Figure 2** – Rietveld refined neutron diffraction pattern for BCY at room temperature: observed, calculated and difference profile.

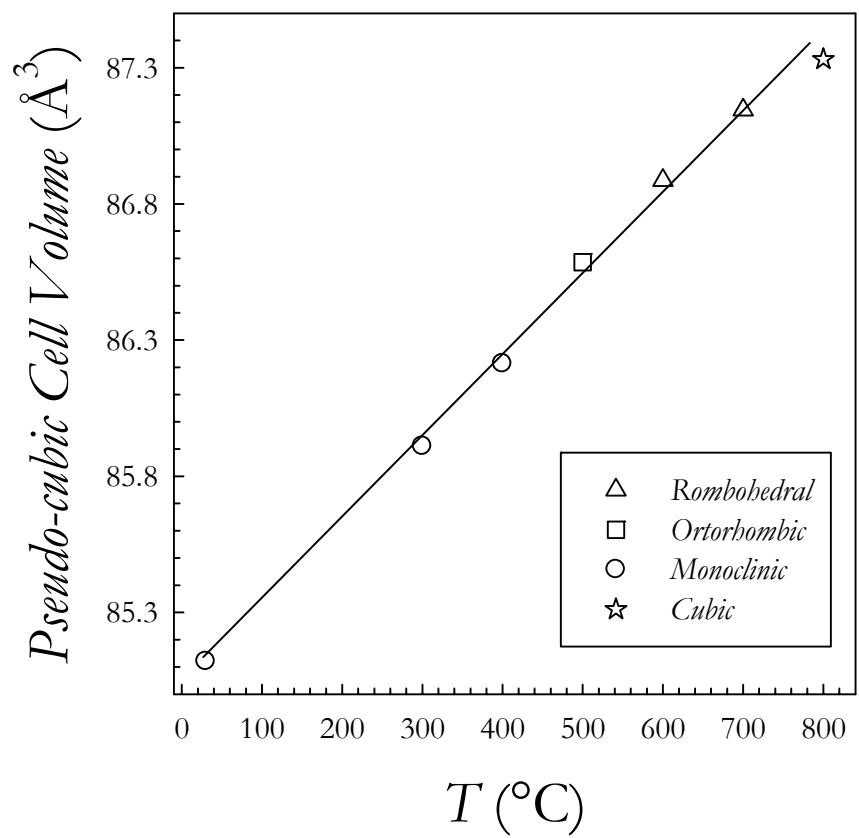

**Figure 3** – Pseudo-cubic cell volume as a function of temperature for BCY. Error bars are smaller then symbols.

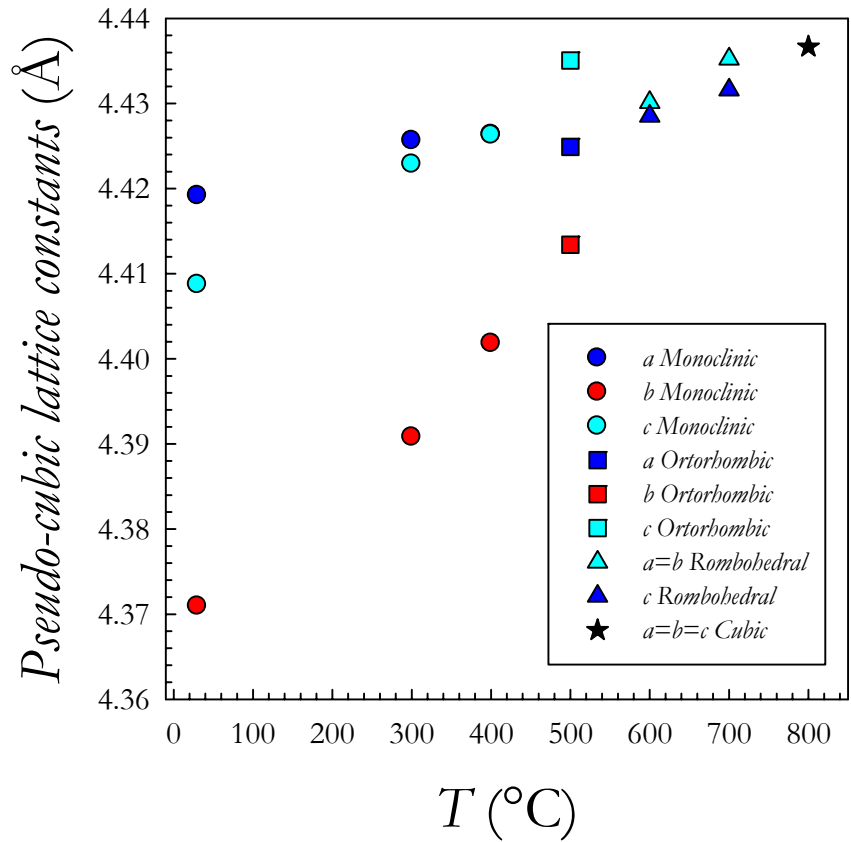

**Figure 4** – Lattice constants as a function of temperature for BCY. Error bars are smaller then symbols. For rhombohedral and orthorhombic samples are reported the pseudo-cubic lattice parameters.

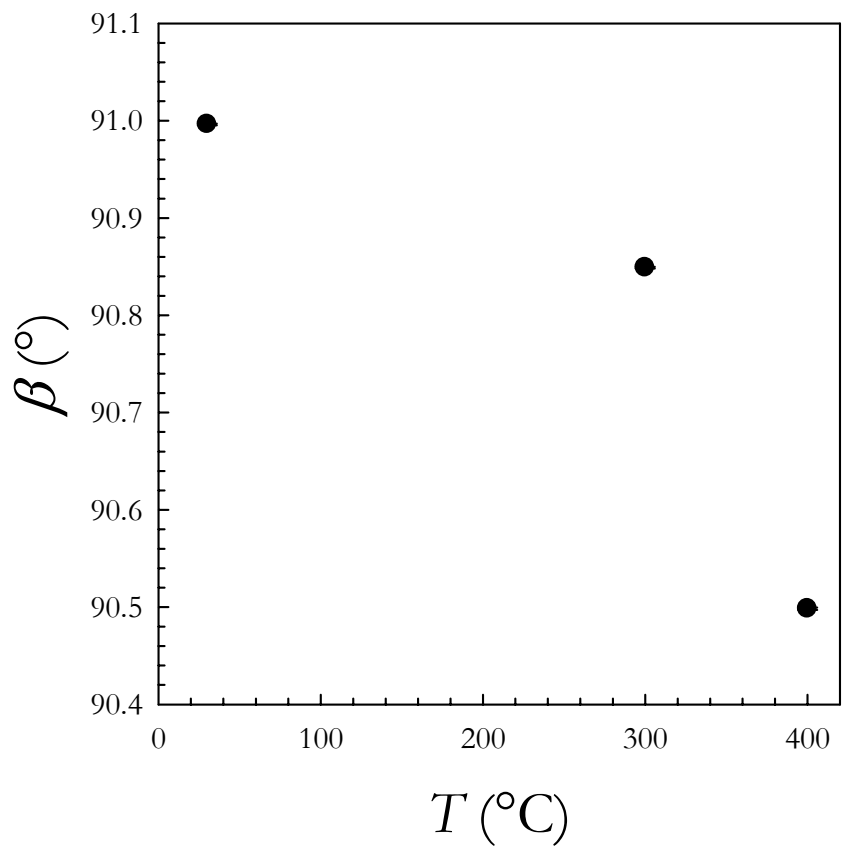

**Figure 5** – Monoclinic *β* angle variation as a function of temperature for BCY. Error bars are smaller then symbols.

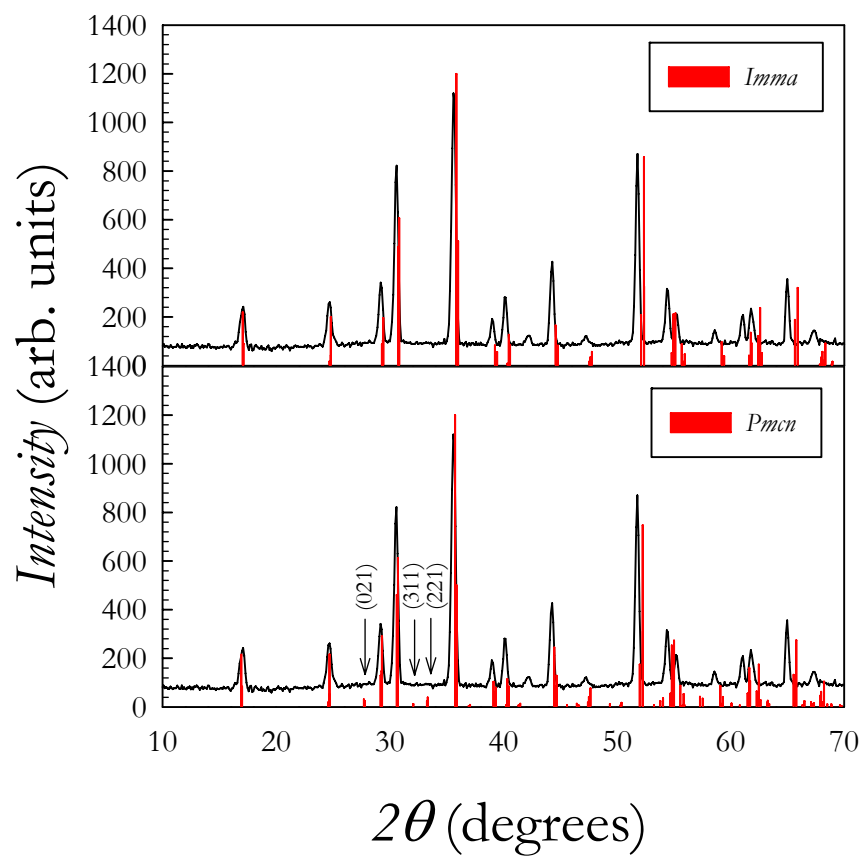

**Figure 6** – Comparison between experimental pattern a 500°C (black line) and calculated diffraction patterns for the *Pmcn* (bottom panel) and *Imma* (top panel) orthorhombic structures.

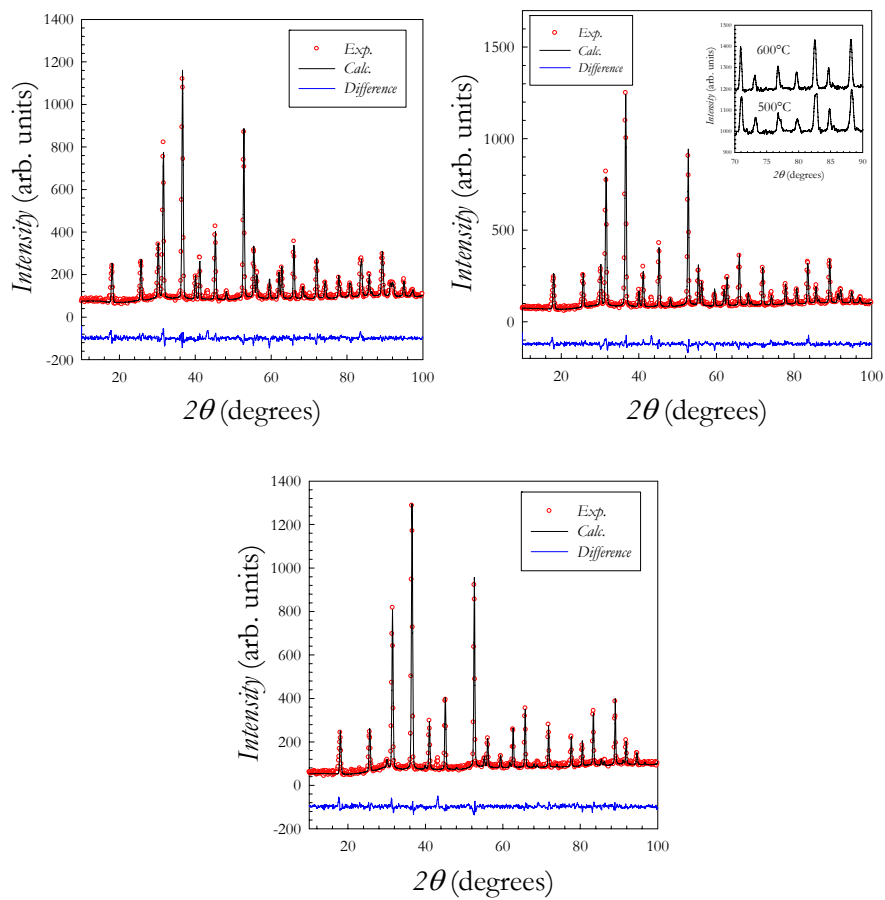

**Figure 7** – a) Rietveld refined neutron diffraction pattern for BCY at 500°C: observed, calculated and difference profile. b) Rietveld refined neutron diffraction pattern for BCY at 600°C: observed, calculated and difference profile. Inset: comparison of a selected region of the ND patterns at 500 and 600°C. c) Rietveld refined neutron diffraction pattern for BCY at 800°C: observed, calculated and difference profile.

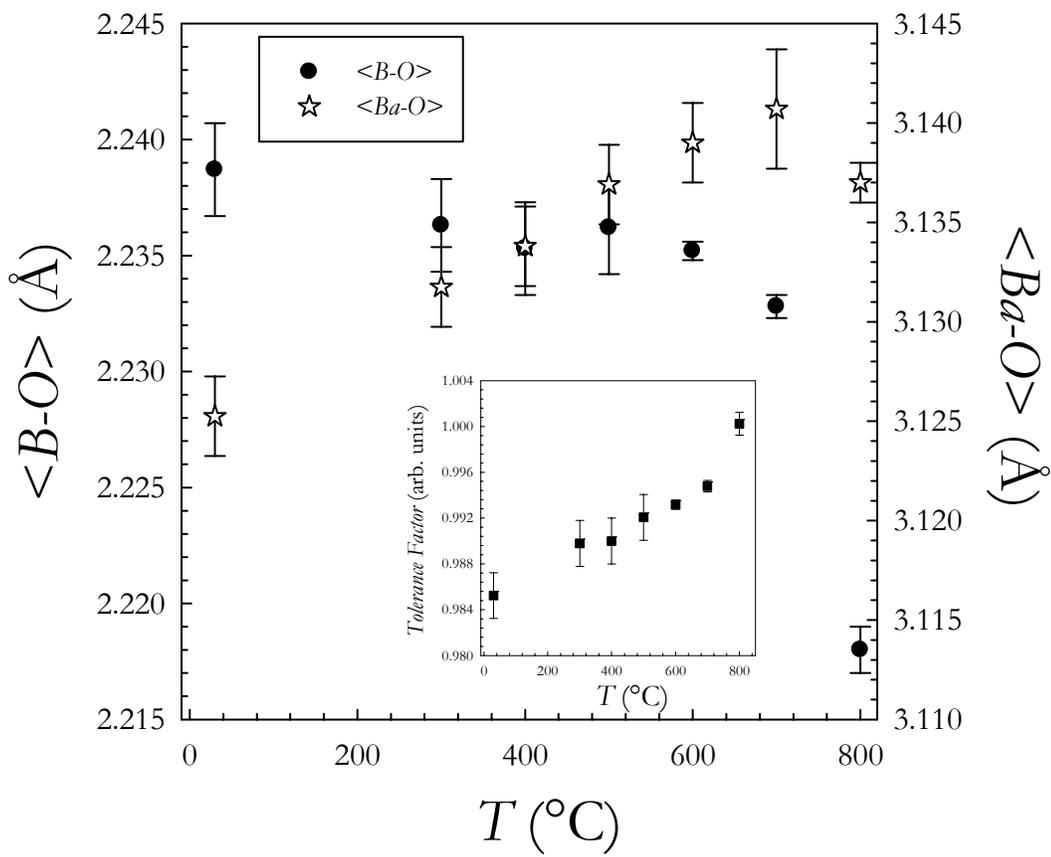

**Figure 8** – Average B-O (full circles) and Ba-O (empty stars) bond lengths evolution with temperature for BCY. Inset: Tolerance factor variation with $T$.

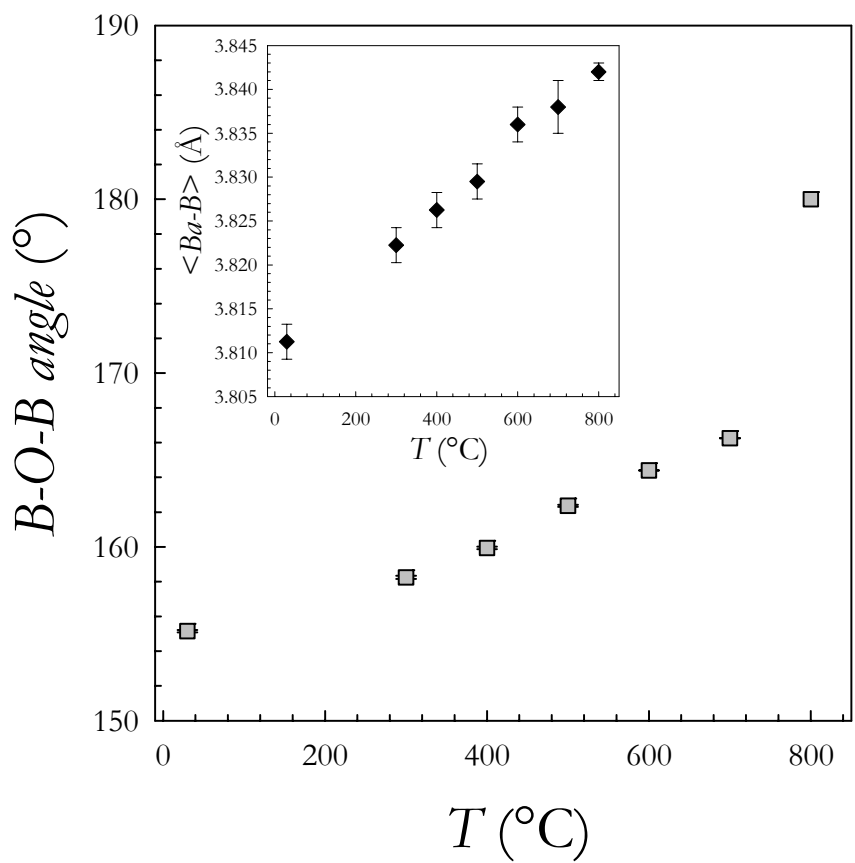

**Figure 9** – Evolution of the average B-O-B bond angle with temperature for BCY. Inset: <Ba-B> bond length variation as a function of temperature.

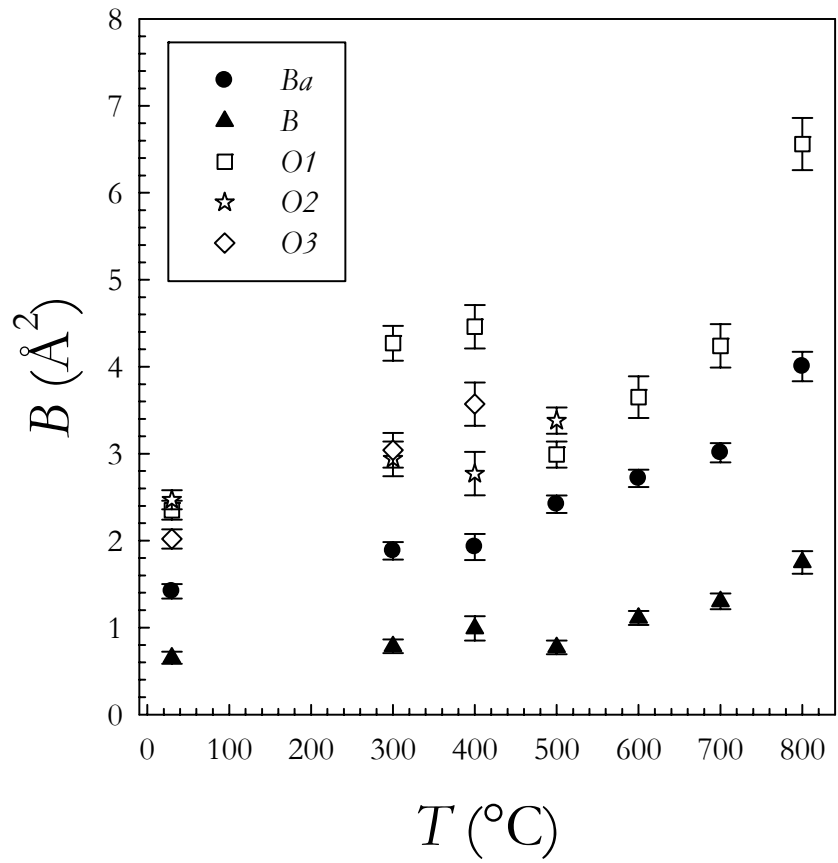

**Figure 10** – Evolution of the atomic displacement parameters for BCY compound as a function of temperature.

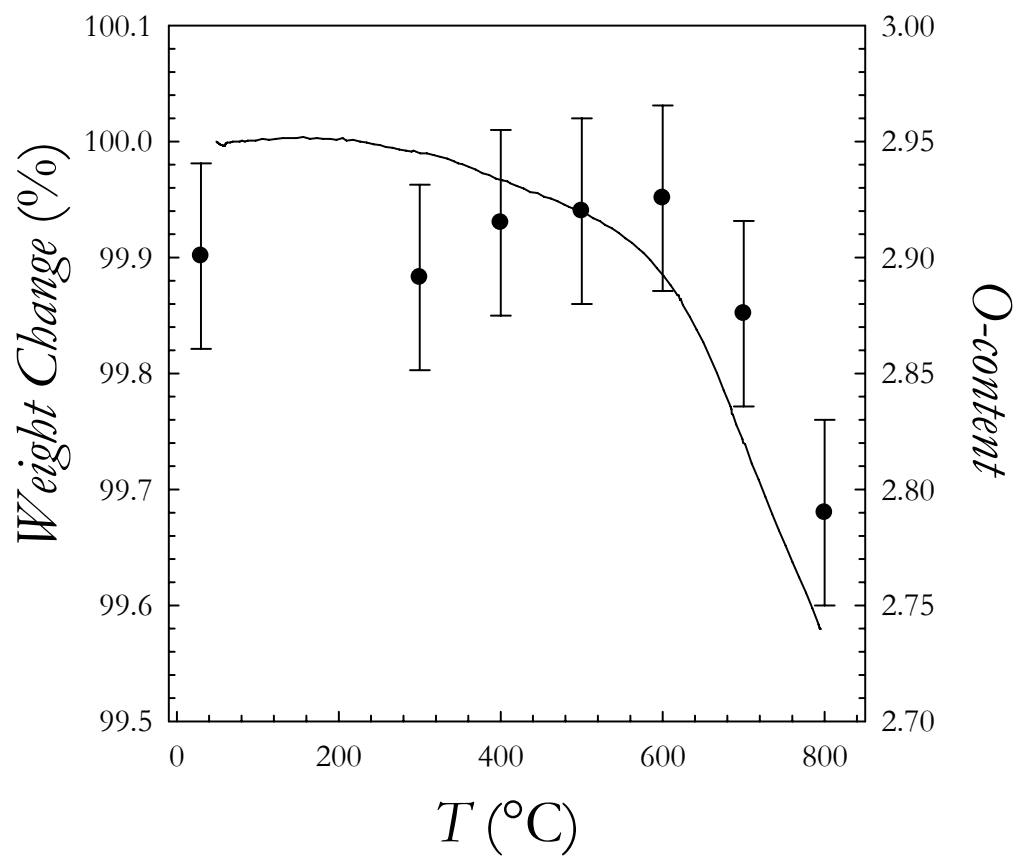

**Figure 11** – TGA trace of BaCe$_{0.80}$Y$_{0.20}$O$_{2.9}$ sample under air (solid line) and oxygen content as a function of temperature as determined from the refinement of the oxygen occupancy.

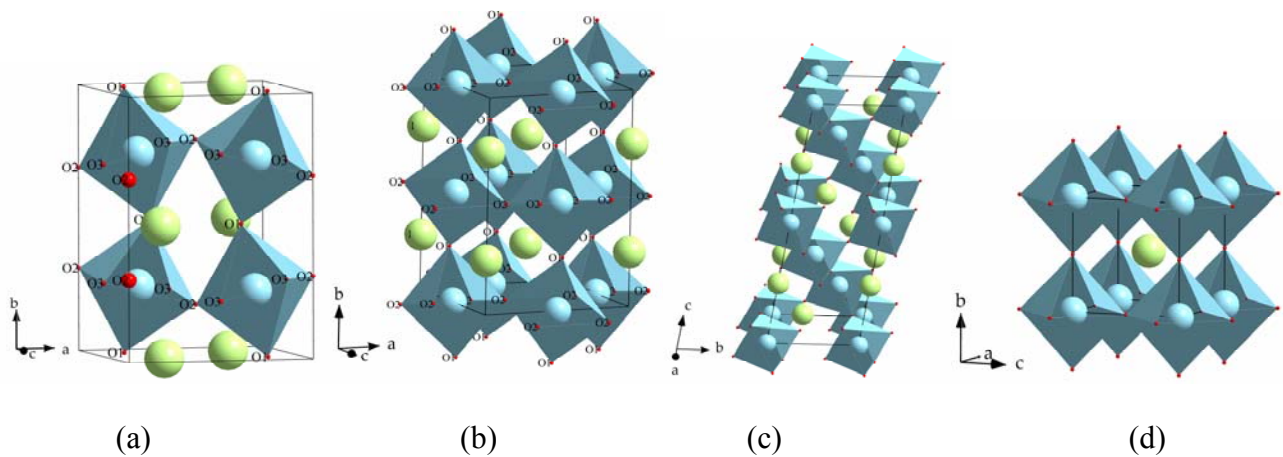

(a)            (b)            (c)            (d)

**Figure 12** – Sketch of the $BaCe_{0.80}Y_{0.20}O_{2.9}$ crystal structure as a function of temperature with B-O polyhedra. Green spheres: Ba ions; blue spheres: Y/Ce ions; Red spheres: oxygen ions. Where more than one O site was present in the structure the oxygen sites were labelled according to the corresponding crystal structure.

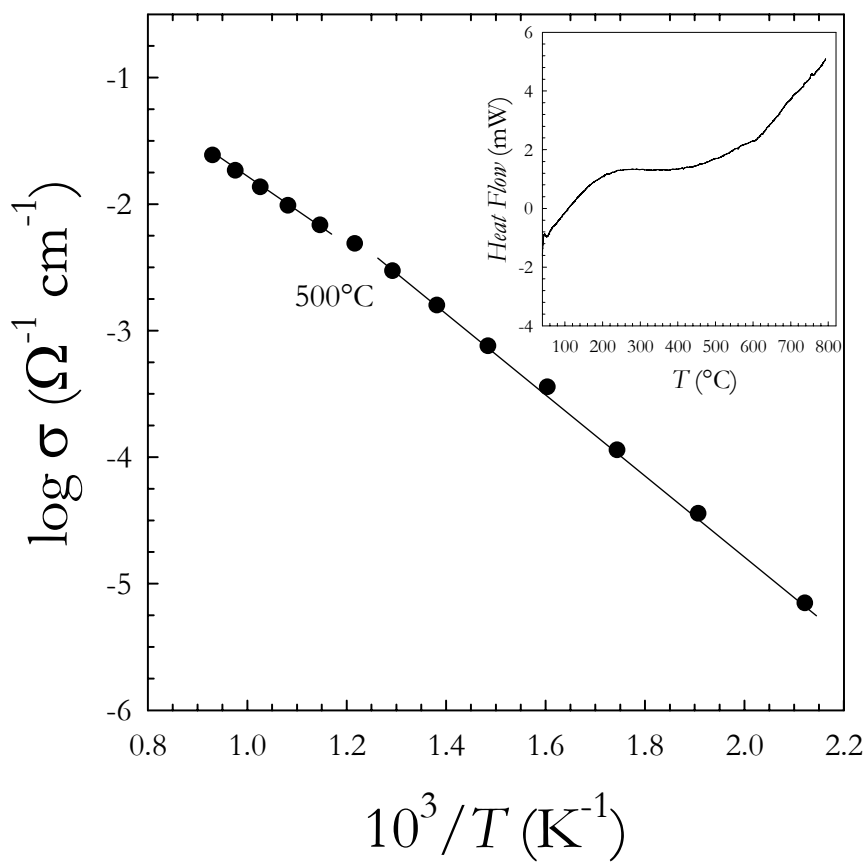

**Figure 13** – Arrhenius plot of electrical conductivity in air for $BaCe_{0.80}Y_{0.20}O_{2.9}$. Inset: DSC trace of $BaCe_{0.80}Y_{0.20}O_{2.9}$.